# LOOPHOLE-FREE BELL TESTS AND THE FALSIFICATION OF LOCAL REALISM

Patrick Fraser[1] & Barry Sanders[2,3,4]

1. Stratford Central Secondary School, 60 St. Andrew St., Stratford ON, N5A 1A3
2. Institute for Quantum Science and Technology, University of Calgary, Alberta T2N 1N4
3. Program in Quantum Information Science, Canadian Institute for Advanced Research, Toronto, Ontario M5G 1Z8, Canada

## ABSTRACT

Quantum mechanics is strictly incompatible with local realism. It has been shown by Bell and others that it is possible, in principle, to experimentally differentiate between local realism and quantum mechanics. Numerous experiments have attempted to falsify local realism; however, they have consistently failed to close the detection loophole under strict locality conditions, thereby allowing local realistic explanations for their observations. In 2015, three experiments took place that tested local realism without the impediments of these significant loopholes. Between these three experiments, a substantial data set was collected. All of the collected data show a strong violation of local realism and strong support for quantum mechanics. This article reviews the theoretical basis of Bell tests and the affiliated loopholes, as well as the methods employed by these recent experiments and the implications of the results they observed.

La mécanique quantique est strictement incompatible avec le réalisme local. Bell et d'autres scientifiques ont montré qu'il est possible, en théorie, de trouver la différence entre le réalisme local et la mécanique quantique expérimentalement. De nombreuses expériences ont tenté de falsifier le réalisme local; cependant, elles ont toujours échoué à combler la faille de détection dans des conditions de localité strictes, permettant ainsi des explications réalistes locales pour leurs observations. En 2015, trois expériences ont testé le réalisme local sans les entraves de ces failles importantes. Entre ces trois expériences, des données substantielles ont été recueillies. Toutes les données recueillies ont montré une forte déviation du réalisme local et un appui solide pour la mécanique quantique. Cet article examine les bases théoriques des tests de Bell et les failles affiliées, ainsi que les méthodes employées par ces expériences récentes et les implications des résultats des expériences.

## INTRODUCTION

Local realism imposes a universal restriction that the properties of all physical objects must always have definite values, thus defining their entire state, and they must only be able to interact with objects within the spatiotemporal limitations of special relativity (Einstein et al., 1935) (Bell, 1964) (Bell, 1976). Within the quantum mechanical description of physical reality, it is possible to have two quantum objects entangled where neither object can be described independently from the other, they can only be considered as a system (Einstein et al., 1935). Measurements on the states of entangled quantum objects (such as photon pairs, electron pairs, etc.) necessarily yield correlated outcomes even when there is sufficient spatiotemporal separation that no sub-luminal signal can mediate the observed correlation (Einstein et al., 1935). Also according to Bohr and Heisenberg, complementary physical quantities such as position and momentum cannot both have definite values simultaneously; their values are probabilistic within the quantum mechanical framework (Bohr, 1935) (Heisenberg, 1927). This means that the physical properties of a quantum object cannot all be fully defined at any given moment. Therefore, local realism is violated by quantum mechanics (Einstein et al., 1935) (Bell, 1964) (Bell, 1976). Einstein, Podolsky, and Rosen argued that this violation of local realism shows that quantum mechanics cannot be considered complete because they assumed that local realism is a necessary condition of physical reality (Einstein et al., 1935). However, Bohr argued that this assumption of local realism is not necessary and that quantum mechanics could be considered complete (Bohr, 1935).

To resolve this, Bell proposed a method of differentiating between local realism and nonlocal theories such as quantum mechanics (Bell, 1964). Bell's theorem is based on the fact that, under quantum mechanics, measurements on entangled objects prepared in the same way will always show correlated results irrespective of spatiotemporal separation (Einstein et al., 1935). For this measured correlation to be observed under local realism, objects must carry information with them about the state they would take on upon measurement (Bohm, 1952).

Bell showed that, if two entangled objects are prepared independently from one another using random analyzer settings, the measured correlation under quantum mechanics would be stronger than that of any local realistic theory (Bell, 1964). Thus it is possible to test local realism experimentally.

An experimentally realizable model that is reducible to Bell's theorem, was proposed by Clauser, Horne, Shimony, and Holt (CHSH) (Clauser et al., 1969). This inequality is subject to the fair sampling hypothesis; since it is impossible to have 100% efficiency experimentally, any experiment testing the CHSH inequality must assume that the detections observed are statistically representative of nature (Clauser et al., 1969). Eberhard further developed this model to determine the minimum detector efficiencies necessary given certain background levels to test the CHSH inequality (Eberhard, 1993). Assuming zero background levels, the absolute minimum efficiency required is $\eta \geq 2/3$ (Eberhard, 1993).

Clauser and Horne (CH) formulated another inequality (Clauser and Horne, 1974). The CH model is not restricted by the assumption that every particle in every entangled pair created is detected, allowing for observed 'singles' to be managed within their framework. This allows an experiment testing the CH inequality to bypass the detection loophole entirely, as discussed below.

Since Bell first showed that it is possible to compare local realism and quantum mechanics using empirical methods rather than strictly philosophical methods, many experiments have sought to do by measuring statistical correlations that cannot be recreated by a local realistic theory (Clauser and Freedman, 1972) (Kasday et al., 1975) (Aspect et al., 1982) (Tittel et al., 1998) (Weihs et al., 1998) (Pan et al., 2000) (Rowe et al., 2001) (Salart et al., 2008) (Giustina et al., 2013) (Christensen et al., 2013). However, none of these experiments have been sufficiently rigorous to falsify local realism because their methods have left room for local realistic explanations due to the detection loophole and the locality condition.

In 2015, three experiments were performed that resolved the most substantial experimental problems known, the locality condition and the detection loophole, making it implausible that their observations could be attributed to some local realistic phenomena (Shalm et al., 2015) (Hensen et al., 2015) (Giustina et al., 2015). Although it is impossible to falsify local realism completely, these experiments have provided compelling, statistically significant evidence in support of quantum mechanics and a violation of local realism (Shalm et al., 2015) (Hensen et al., 2015) (Giustina et al., 2015).

## CRITERIA FOR A LOOPHOLE-FREE BELL TEST

To falsify local realism, an experiment must observe phenomena that could not be explained by local realism. This can be accomplished by measuring the statistical correlation between the states of entangled particles and comparing them to the expected values of Bell-type inequalities (Bell, 1964).

In a typical Bell test, entangled pairs are generated by some source (Einstein et al., 1935) (Bell, 1964). Both entangled objects then diverge from one another (Einstein et al., 1935) (Bell, 1964). Each object is prepared by an analyzer on its own 'side' before being detected (Einstein et al., 1935) (Bell, 1964). A test statistic (often denoted $S$ for CHSH experiments) is usually reported, representing the statistical correlations between the number of detection coincidences under certain analyzer settings (Clauser et al., 1969). A higher test statistic value indicates a greater correlation between the measurements on both sides of the experiment.

The value $S$ is defined for a given CHSH experiment by

$$S = E(a, b) - E(a, b') + E(a', b) + E(a', b')$$

where $a$ and $a'$ are analyzer settings on side $A$, $b$ and $b'$ are analyzer settings on side $B$, and $E(a, b)$ is defined as

$$E(a, b) = \frac{N_{++} - N_{+-} - N_{-+} + N_{--}}{N_{++} + N_{+-} + N_{-+} + N_{--}}$$

where $N_{++}$ is the number of coincidences where both particles are detected, $N_{+-}$ and $N_{-+}$ are the number of coincidences where only one particle is measured, and $N_{--}$ is the number of coincidences where neither particle is observed (Clauser et al., 1969). By randomly preparing two particles and randomly measuring each in a binary valued, two setting experiment, the correlations of the measurements are bound by different inequalities for local realistic and nonlocal theories (Clauser et al., 1969). Under local realism, the test statistic $S$ for a CHSH test is bound by $|S| \leq 2$ (Cirel, 1980). However, in nonlocal theories such as quantum mechanics, this value has an upper bound of $|S| \leq 2\sqrt{2}$ (Cirel, 1980) which is a mathematical consequence of the inseparability of entangled states and the noncommutativity of operators.

The test statistic for a CH experiment is

$$-1 \leq p_{12}(a,b) - p_{12}(a,b') + p_{12}(a',b) + p_{12}(a',b') - p_1(a') - p_2(b) \leq 0$$

where $p_{12}(a,b)$ is the probability that both particles will be detected under analyzer settings $a$ and $b$ (Clauser and Horne, 1974).

However, these inequalities are theoretical and practical experiments are always prone to imperfections. In order for the results to falsify local realism and be considered *strictly loophole-free*, no local realistic explanation may exist for the experimentally observed phenomena. For example, if entangled particles are measured near one another, it may be possible for some locally acting signal to communicate between the particles and 'tell' them whether to pass through an analyzer or not as such a signal could travel slower than light. Although there is no indication of any such signal existing, it cannot be ruled out as a possibility unless the detectors have sufficient spatiotemporal separation to make any local communication impossible (Bell, 1964).

A local realistic explanation may include any theoretical mechanism or model of the universe that can make the results of a given experiment appear nonlocal (Horodecki et al., 2014) (Koh et al., 2012). For a Bell test to be *experimentally loophole-free* it must only meet the condition of locality and close all loopholes that can, in principle, be closed experimentally, primarily the detection loophole.

# REQUIREMENT OF RANDOMNESS

For a given Bell test, to ensure that any local realist theory has no information about the experiment analyzer settings in advance, random analyzer settings must be generated and thus random number generators (RNGs) are required (Hall, 2010). However, in classical physics, which maintains the assumption of local realism, true randomness is not possible as every interaction is strictly deterministic (Laplace, 1902). Even the best computer-based RNGs act algorithmically and are thus still deterministic and can only be considered pseudo-random, requiring seeds, which may themselves be correlated (Weihs et al. 1998).

Within the framework of quantum mechanics, it is possible to generate true randomness (Pironio et al., 2010) (Colbeck and Renner, 2011) (Frauchiger et al., 2013). However, in any experiment testing Bell's inequality or a variant of it, the aim is to measure statistical results that either falsify or corroborate local realism (Bell, 1964). By assuming that local realism is true, quantum mechanical processes, which fundamentally require violations of local realism, may not be employed. Otherwise, the experiment would be methodologically dependent on the falsity of local realism; this is a fallacious petitio principii argument.

Since quantum mechanics presents the only known source of true randomness, only pseudo-random number generators (PRNGs) can be used in Bell test experiments. Therefore, local realism can never be fully falsified since doing so would require true randomness which is fundamentally impossible to produce under the deterministic framework of local realism (Brans, 1988). Thus, it is impossible, even in principle, to carry out a *strictly loophole-free* Bell test (Larsson, 2014).

For Bell test experiments, since only PRNGs can be used, the probability $p$ that the pseudo-random analyzer settings could be predicted and thus used by a local realistic theory can be computed; lesser $p$ values imply more random sources (Barrett, 2002).

## BELL TEST LOOPHOLES

Many experiments have tested the relationship between entangled particles and have shown violations of Bell's inequality and the CHSH inequality however, all of these experiments failed to close all experimentally avoidable loopholes (Clauser and Freedman, 1972) (Kasday et al., 1975) (Aspect et al., 1982) (Tittel et al., 1998) (Weihs et al., 1998) (Pan et al., 2000) (Rowe et al., 2001) (Salart et al., 2008) (Giustina et al., 2013) (Christensen et al., 2013).

## THE LOCALITY CONDITION

For many Bell test experiments, the spatiotemporal separation of the analyzers was less than that required for the interactions to be nonlocal (Freedman and Clauser, 1972) (Kasday et al., 1975) (Aspect et al., 1982) (Tittel et al., 1998) (Pan et al., 2000) (Rowe et al., 2001) (Ansmann et al., 2009) (Christensen et al., 2013) (Giustina et al., 2013). No phenomena observed experimentally can be nonlocal if the distance between analyzers $d$ and the detection times $t$ are such that $d/t \leq c$ as they could then be reasonably explained within the restrictions of special relativity (Bell, 1964). In order for an experiment to truly test local realism, it is necessary that any interactions restricted by special relativity are made impossible (Bell, 1964).

To date several experiments have been performed that satisfied this locality condition (Salart et al., 2008) (Weihs et al., 1998) (Giustina et al., 2015) (Shalm et al., 2015) (Hensen et al., 2015). In each of these experiments, the results showed a definite violation of some Bell-type inequality in agreement with the predictions of quantum mechanics (Salart et al., 2008) (Weihs et al., 1998) (Giustina et al., 2015) (Shalm et al., 2015) (Hensen et al., 2015). However, several of these experiments were still subject to other loopholes, primarily the detection and memory loopholes (Salart et al., 2008) (Weihs et al., 1998).

## THE DETECTION LOOPHOLE

As Pearle showed, Bell's inequality and the CHSH inequality assume that all possible outcomes for both particles in a two-sided Bell test experiment can be measured (Pearle, 1970). However, in many cases, one or both of the particles will go undetected (Pearle, 1970). Thus, it must be assumed that the coincidences measured are statistically representative of nature (Pearle, 1970). This fair sampling hypothesis allows for the existence of local realistic theories that prevent certain events from being detected (Pearle, 1970) (Clause and Horne, 1974). In this case, apparently nonlocal observations may be explained within local realism (Pearle, 1970).

To close this loophole, the number of unobserved events must be sufficiently low that they could not lead to a local realistic explanation (Pearle, 1970). Otherwise, an experiment must test the CH model, which excludes unobservable events from its derivation (Clauser and Horne, 1974).

The first experiment to close the detection loophole had a detector efficiency greater than 90% thereby ensuring that the results were indicative of a violation of the CHSH inequality (Rowe et al., 2001). However, this experiment was not performed under strict locality conditions. Therefore, the results still allow for a local realistic explanation as it failed to meet the locality condition (Rowe et al., 2001). Other experiments have also been performed that closed the detection loophole (Ansmann et al., 2009) (Christensen et al., 2013) (Giustina et al., 2013). However, none of these tests satisfied the locality condition (Christensen et al., 2013) (Giustina et al., 2013) (Ansmann et al., 2009).

## THE MEMORY LOOPHOLE

In Bell tests, if the analyzer settings go unchanged during several trials, the analyzers can interact locally with the point where the entangled pairs are formed over many trials (Bell, 1964) (Aspect, 1982). Thus, there is the possibility for a local realistic theory which uses the known information about analyzer settings to influence the states of the entangled pairs as they are created (Barrett et al., 2002). Therefore, if the analyzers remain fixed during several trials of an experiment, it is possible under local realism for the results to appear nonlocal (Bell, 1964) (Aspect et al., 1982). In order to bypass this problem, analyzer settings must vary in time, as was first accomplished by Aspect et al. (Aspect et al.,1982).

Furthermore, if the analyzers vary in some fixed pattern, it is still theoretically possible for the predictability of the settings to be used to influence the entangled states (Aspect et al.,1982) (Weihs et al., 1998) (Horodecki et al., 2014). In order to close the memory loophole, settings for analyzers on either side must be generated randomly (Hall, 2010). These settings must be determined by RNGs that are sufficiently distant and fast that separate random numbers can be generated after the entangled pair is created but before the entangled particles arrive at the analyzers (Bell, 1964) (Aspect, 1982). This must also be done without the possibility of the RNGs interacting locally with one another as they could then become correlated (Brans, 1988).

The memory loophole can never be totally closed by any practical experiment; closing it would require the use of truly random number generation (Brans, 1988) (Hall, 2010). Within the context of a Bell test experiment, this is not allowable since Bell tests must assume a deterministic framework (Brans, 1988). The closest an experiment can come to closing the memory loophole is to use PRNGs that are as unpredictable as possible, thereby maximizing the restrictions on local realistic theories (Brans, 1988).

## DISCUSSION

Although many experiments have been performed seeking to falsify local realism, they have all had one or more loopholes allowing for local realistic explanations (Wu and Shaknov, 1950) (Clauser and Freedman, 1972) (Kasday et al., 1975) (Aspect et al., 1982) (Tittel et al., 1998) (Weihs et al., 1998) (Pan et al., 2000) (Rowe et al., 2001) (Salart et al., 2008) (Ansmann et al., 2009) (Giustina et al., 2013) (Christensen et al., 2013). This remained the case until the first loophole-free Bell tests were performed in 2015 (Giustina et al., 2015) (Shalm et al., 2015) (Hensen et al., 2015).

*Bell Inequality Violation Using Electron Spins*

In 2015, Hensen et al. became the first group to successfully carry out a loophole-free Bell test experiment (Hensen et al., 2015). Their procedure involved creating pairs of entangled electrons that were located 1.3 km apart (Hensen et al., 2015). To do this, they isolated electrons in nitrogen vacancy defect centres in synthetic diamond chips, allowing them to easily manipulate the electron spins (Hensen et al., 2015). These electrons were then excited using microwave radiation, each emitting a single photon (Hensen et al., 2015). Once both electrons had emitted a photon, the photons travelled via fiber optic cables to a central location where they interacted with one another (Hensen et al., 2015). Since the individual electrons were entangled with the photons they emitted, and the photons became entangled with each other following their interaction, the two electrons became entangled (Hensen et al., 2015).

By measuring the spin of the electrons independently and following the photon interactions in a sufficiently small period of time, any observed correlation could not be caused by a sub-luminal signal (Hensen et al., 2015). Therefore, this experiment successfully met the locality condition (Hensen et al., 2015).

By performing a Bell test that tested the CHSH model, and by using detectors with sufficient efficiency to satisfy the requirements put forth by Eberhard thus making the 'singles' rate negligible, this experiment successfully closed the detection loophole (Eberhard, 1993) (Hensen et al., 2015). By using fast-changing, pseudo-random analyzer settings, this experiment was able to reduce the predictability of analyzer settings to $p = 0.039$ (Hensen et al., 2015). This mitigated the impact of the unavoidable memory loophole. Therefore, the experiment performed by Hansen et al. closed the significant, experimentally avoidable loopholes and was thus experimentally loophole-free (Hensen et al., 2015).

This experiment consisted of 245 trials (Hensen et al., 2015). The results showed a test statistic value of $S = 2.24 \pm 0.20$ (Hensen et al., 2015). Therefore, this experiment violated the requirements of local realism while staying within the limits of quantum mechanics (Hensen et al., 2015). They concluded that these results implied a rejection of the local realistic null hypothesis (Hensen et al., 2015).

*Strong Loophole-Free Test of Local Realism*

In the experiment performed by Shalm et al., pairs of polarization-entangled photons were generated by means of spontaneous parametric down conversion; a process wherein a high-frequency photon enters a nonlinear crystal, such as BBO or PPKTP, and is spontaneously converted into two entangled photons of lower energy (Burnham and Weinberg, 1970) (Shalm et al., 2015). These entangled photons then passed through fiber optic cables to locations of detection with a sufficiently large spatiotemporal separation to meet the locality condition (Shalm et al., 2015).

Analyzer settings for each trial were determined by performing an XOR operation on three pseudo-randomly generated bits (Shalm et al., 2015). These bits were determined by sequences from popular culture movies and TV shows as well as strings of digits from π to create pseudo-random bits. Shalm et al. claimed that any local realistic theory capable of predicting these values in advance would need to have predictability such that it could create cultural artifacts and thus have elements of superdeterminism (Shalm et al., 2015).

Furthermore, the PRNGs had sufficient spatiotemporal separation to produce analyzer settings while the entangled photons were en route to the analyzers (Shalm et al., 2015). In doing this, Shalm et al. were able to ameliorate the effect of the memory loophole as well (Shalm et al., 2015).

Finally, by using highly efficient detectors and testing a version of the CH model, they were able to close the detection loophole (Shalm et al., 2015). The necessary theoretical efficiency for this experiment, which Shalm et al. calculated using the method proposed by Eberhard, was 72.5% (Eberhard, 1993) (Shalm et al., 2015). The actual detector efficiencies used were $74.7 \pm 0.3\%$ and $75.6 \pm 0.3\%$ as calculated using the method proposed by Klyshko (Klyshko, 1980) (Shalm et al., 2015).

This experiment reported 706,555,817 trials performed for which the adjusted $p$ value was $p = 2.3 \times 10^{-7}$ (Shalm et al., 2015). This implies statistically significant falsification of local realism (Shalm et al., 2015).

The experimenters explicitly stated that this experiment still required the assumptions that their spatiotemporal measurements were accurate, that the measured results were fixed at the noted time taggers, and that pseudo-random number generation was totally independent for both detection systems (Shalm et al., 2015). However, these assumptions can never, even in principle, be removed fully (Shalm et al., 2015). Thus, under the required assumptions, this Bell test closed the most significant, experimentally avoidable loopholes and demonstrated a significant violation of the CH inequality (Shalm et al., 2015). The results could not have been produced by a non-superdeterministic local realistic theory with high probability (Shalm et al., 2015).

*Test of Bell's Theorem with Entangled Photons*

In the experiment carried out by Giustina et al., pairs of entangled photons were generated using high-efficiency spontaneous parametric down conversion (Giustina et al., 2015). Using independent, time synchronized PRNGs, the settings of polarization analyzers were determined while the photons were traveling from the source to the detectors (Giustina et al., 2015). This minimized the ability of any local realistic theory to predict the analyzer settings, thereby minimizing the impact of the memory loophole (Giustina et al., 2015).

The detectors were each located approximately 29 m from the source and approximately 58 m from each other (Giustina et al., 2015). Photon detection required a time interval of 64.4 ns to 65.5 ns (Giustina et al., 2015). Thus, it can be concluded that the detectors were sufficiently spatially separated to prevent any subluminal or luminal communication between them during each trial (Giustina et al., 2015). Therefore, this experiment met the condition of locality (Giustina et al., 2015).

By using a procedure that employed a CH-Eberhard inequality that holds for all local realistic theories

$$J \equiv p_{++}(a,b) - p_{+0}(a,b') + p_{0+}(a',b) - p_{++}(a',b') \leq 0$$

where $J$ represents the correlation between the observations on both sides and $p_{++}(a,b)$ is the probability that, under analyzer settings $a$ and $b$, both particles are detected, the results obtained were not subject to the fair sampling hypothesis (Giustina et al., 2015). This means that this experiment effectively closed the detection loophole as well (Giustina et al., 2015). Since the reported $J$-value of $7.27 \times 10^{-6}$ was greater than zero, this experiment effectively showed a violation of local realism (Giustina et al., 2015).

In their supplemental material, Giustina et al. reported the number of valid trials they performed as $N = 3{,}502{,}784{,}150$ (Giustina et al., 2015). Consequently, they reported a $p$ value of $p = 3.74 \times 10^{-31}$ corresponding to 11.5 standard deviations meaning there is a very low probability that these results could be credited to a local realist explanation (Giustina et al., 2015). Therefore, this experiment shows a statistically significant violation of local realism (Giustina et al., 2015).

*Superdeterminism*

To account for the apparent nonlocality observed in these experiments, either local realism must be false, or the universe must be superdeterministic (Larsson, 2014). This conflict of ideas holds significant meaning for how physicists understand the universe (Hooft, 2007) (Larsson, 2014). Although the notion of superdeterminism is important, it cannot itself be falsified and is therefore not a scientific theory, testable by empirical methods (Larsson, 2014).

The nonlocal interactions apparently observed in loophole-free Bell tests could, in principle, be caused by superdeterminism under local realistic conditions because superdeterminism requires that the outcomes of all interactions were determined at the outset of the universe (Colbeck and Renner, 2012), (Larsson, 2014). Thus, the outcomes of the experiments performed are 'decided' before the experiment begins so no amount of spatial separation can prevent the inevitable (Colbeck and Renner, 2012) (Larsson, 2014).

Since superdeterminism is unfalsifiable, it will forever remain a valid alternative to any empirical scientific theory no matter how well that theory predicts the measured observations (Larsson, 2014). Therefore, superdeterminism can never be discarded but must remain a possibility even when all evidence points towards a quantum mechanical, nonlocal universe (Larsson, 2014).

## CONCLUSION

Based on current knowledge, quantum mechanics and the nonlocality associated with it has been sufficiently corroborated to make it the most accurate physical theory available. With several groundbreaking experiments closing the detection loophole and maintaining strict locality conditions simultaneously, and showing statistically significant violations of the CHSH and CH inequalities, it appears as though the universe acts nonlocally, within the limits of a quantum theory (Giustina et al., 2015) (Shalm et al., 2015) (Hensen et al., 2015).

In addition to these conclusions about the fundamental nature of physical reality, these findings also have significant implications in the practical applications of quantum information (Frauchiger et al., 2013) (Colbeck and Renner, 2011) (Pironio et al., 2010). If local realism is falsified and quantum mechanics is corroborated, assuming the universe does not act superdeterministically, it becomes possible to generate true randomness, allowing secure device-independent quantum cryptographic systems to be made practical (Frauchiger et al., 2013) (Colbeck and Renner, 2011) (Pironio et al., 2010).

The results discussed have shown strong falsification of local realism however, this does necessitate quantum as an explanation (Guistina, 2016). In addition to applying these ideas to practical technologies, further research may be done to further corroborate quantum mechanics as the most accurate nonlocal theory of the universe. Finally, nonlocality should be explored further as it presents new problems regarding causality.

## ABBREVIATIONS

| Abbreviation | Full Form |
| --- | --- |
| CHSH | Clauser-Horne-Shimony-Holt |
| CH | Clauser-Horne |
| RNG | Random Number Generator |
| PRNG | Pseudo-Random Number Generator |

## ACKNOWLEDEMENTS

We are indebted to the Foundation for Student Science and Technology and their student Co-op program for making the opportunity to carry out this work possible. PF also thanks Susan Young and Mary Lou Ricard for their encouragement. BCS appreciates financial support from Alberta Innovates.